\documentclass[a4paper]{article}
\usepackage{indentfirst}
\usepackage{graphicx}
\usepackage{setspace}
   \title{Realizing Tri-bimaximal Mixing in Minimal Seesaw Model with S4 Family Symmetry}
  \author{Zhen-hua Zhao\\Institute of Theoretical Physics, Chinese Academy of Sciences,\\
and State Key Laboratory of Theoretical Physics,\\
P.O. Box 2735, Beijing 100190, China}
   \begin{document}
\maketitle
\begin{abstract}
In this paper, we realize the tri-bimaximal mixing in the lepton sector in the context of minimal seesaw in which only two right-handed neutrinos are introduced, with the discrete group $S4$ as the family symmetry. In order to constrain the form of superpotential, a $Z3$ symmetry is also introduced. In the model, the mass matrices for charged leptons and right-handed neutrinos are diagonal. The unitary matrix that diagonalizes the light Majorana neutrino mass matrix is exact tri-bimaximal at LO, and is corrected by small quantities of $\mathcal{O}(0.01)$ at NLO. The mechanism to get the particular scalar VEV alignments used is also presented. Phenomenologically, the mass spectrum is of normal hierarchy with $m_1=0$, and $\sum m_i$ and $|m_{ee}|$ are about $0.058$ $eV$ and $0.003$ $eV$ respectively.

\end{abstract}
 \setlength{\parskip}{0.5\baselineskip}

\section{Introduction}
Nowadays, that neutrinos oscillate among three active neutrinos has been established as a fact by many experiments ranging from solar and atmospheric neutrino experiments to reactor and accelerator neutrino experiments \cite{1}. Thanks to the accumulation of data, the mixing angles and mass-squared differences that describe neutrino oscillations now can be determined with a high precision. According to a global neutrino oscillation data analysis within the three-flavor framework \cite{2}, the best fit values of oscillating parameters are given as following, $\Delta m^2_{12}=(7.59^{+0.20}_{-0.18})\times 10^{-5} eV^2$, $\Delta m^2_{13}=(2.45^{+0.09}_{-0.09})\times 10^{-3} eV^2$ and $\sin^2(\theta_{12} ) =0.312^{+0.017}_{-0.015}$, $\sin^2(\theta_{23} ) =0.51^{+0.06}_{-0.06}$, $\sin^2(\theta_{13} )=0.010^{+0.009}_{-0.006}$.

Remarkably, the mixing angles are highly consistent with that of the so-called tri-bimaximal mixing matrix proposed by Harrison, Perkins and Scott \cite{3},
\begin{equation}
     U_{TB}=\left(\begin{array}{ccc}
       \frac{2}{\sqrt{6}} &\frac{1}{\sqrt{3}}&0\\
       \frac{-1}{\sqrt{6}}&\frac{1}{\sqrt{3}}&\frac{-1}{\sqrt{2}}\\
       \frac{-1}{\sqrt{6}}&\frac{1}{\sqrt{3}}&\frac{1}{\sqrt{2}}
\end{array}
\right),
\end{equation}
which consists of $\sin^2(\theta_{12} ) =\frac{1}{3}$, $\sin^2(\theta_{23} ) =\frac{1}{2}$, $\sin^2(\theta_{13} )=0$. The particular oscillating pattern possessed by neutrinos might be just an accident, however, if taken seriously, it maybe imply an underlying family symmetry in the lepton sector. Based on this point of view, many discrete groups \cite{4}, one of which is $S4$ \cite{5}, have been considered as the family symmetry to study Fermion mass matrices.

On the other hand, the smallness of neutrino masses is explained by the well-known seesaw mechanism \cite {6} which can realize the dimension-five Weinberg's operator\cite{7} by introducing heavy particles in addition to the SM particles. The simplest way to realize the seesaw mechanism which is compatible with observational neutrino mass spectrum is to introduce two heavy right-handed neutrinos and this case is called the minimal seesaw. Ever since the celebrated paper by Frampton, Glashow and Yanagida \cite{8}, minimal seesaw models assuming one or two zeros in the Dirac mass matrix which can lead to predictive results due to less parameters have been extensively studied \cite{9,10}. Interestingly, in \cite{11}, the authors make a bottom-up analysis and derive what form the Dirac mass matrix should have so that the mixing matrix is tri-bimaximal in the light neutrino sector after seesaw mechanism is carried out.

In this paper, we discuss a model which can naturally give the tri-bimaximal mixing in the lepton sector in the context of minimal seesaw, using $S4$ as the family symmetry. In \cite{11}, all the special forms of the $3 \times 2$ Dirac mass matrix in the neutrino sector which can lead to the tri-bimaximal mixing are given, when the mass matrices for charged leptons and right-handed neutrinos are diagonal. Our model makes one form of those listed in \cite{11} come out naturally from an underlying theory which has the $S4$ family symmetry. In the model, the mass spectrum is of normal hierarchy with $m_1=0$. The mixing matrix is exact tri-bimaximal at LO, and is corrected by small quantities at NLO. In addition, the model is predictive because there is less effective parameters in the mass matrix than conventional seesaw model in which three right-handed neutrinos are introduced.

We notice that in \cite{12} and \cite{13}, the authors also discuss how to get the tri-bimaximal mixing with $S4$ flavor symmetry in the minimal seesaw model. However, their models are different from ours in field contents, charge assignments and the phenomenological consequences. Besides, in \cite{12} the authors assume some particular relations among the VEVs of different Higgs fields and different Yukawa coupling coefficients without giving the underlying theory while in our model the particular forms possessed by the mass matrices come out naturally once suitable particle contents and charge assignments are chosen; in \cite{13} the authors assume that the VEVs of scalar fields have some particular forms but do not discuss the scalar potential in detail while in this paper we will examine the scalar potential and show that the particular VEV alignment is indeed the minimum.

The other parts of this paper are organized as follows: We first simply review some fundamental aspects of $S4$ which will be used throughout all the paper in Section 2. The model which can naturally realize the tri-bimaximal mixing in the lepton sector will be presented in Section 3 and some phenomenological consequences will also be discussed. In Section 4, we show that the particular VEV alignment which is essential to realizing the tri-bimaximal mixing can come out naturally due to the special form of the scalar potential. In Section 5, we discuss the impact of NLO corrections. Finally, the results are summarized in Conclusions.

\section{$S4$ revisited}
$S4$, the permutation group of 4 objects, consists of 24 elements and 5 irreducible representations which are denoted as $1_1$, $1_2$, $2$, $3_1$, and $3_2$  where the first number indicates dimension of the representation while the subindex is used to distinguish inequivalent representations of the same dimension. The presentation of $S4$ we will use is the one reported in Appendix A of \cite{14}. In the literature, different presentations \cite{15,16} have also been considered and are related to each other by unitary transformations which will not affect the physics. This chosen presentation is suitable in the sense that the mass matrices of charged leptons and right-handed neutrinos are both diagonal as we shall see which will bring us with much convenience and make the tri-bimaximal mixing more transparent. We only list the multiplication rules below and refer the reader to \cite{14} for more details about the Clebsch-Gordan coefficients.

\setlength{\parskip}{0.5\baselineskip}
$1_1$ $\bigotimes$ $\eta$ \  = $\eta$ \hspace{2cm}with $\eta$ any representation

$1_2$ $\bigotimes$ $1_2$  = $1_1$

$1_2$ $\bigotimes$ $2$ \    = $2$

$1_2$ $\bigotimes$ $3_1$  = $3_2$

$1_2$ $\bigotimes$ $3_2$  = $3_1$

$2$\ \ $\bigotimes$ $2$  \    = $1_1$ $\bigoplus$ $1_2$ $\bigoplus$ $2$

 $2$\ \    $\bigotimes$ $3_1$   = $3_1$ $\bigoplus$ $3_2$

 $2$\ \   $\bigotimes$ $3_2$   = $3_1$ $\bigoplus$ $3_2$

$3_1$ $\bigotimes$ $3_1$  = $3_2$ $\bigotimes$ $3_2$ = $1_1$ $\bigoplus$ $2$ $\bigoplus$ $3_1$ $\bigoplus$ $3_2$

 $3_1$ $\bigotimes$ $3_2$ = $1_2$ $\bigoplus$ $2$ $\bigoplus$$3_1$ $\bigoplus$ $3_2$.

\section{The structure of the model\\ and phenomenological analysis}

 \setlength{\parskip}{0.5\baselineskip}

    \begin{table}[tbp]
\centering
\begin{tabular}{lcccccccccccccc}
\hline
 &$L$ & $e^c$ &($\mu^c$,$\tau^c$)& $N_1$ & $N_2$ & $h_{u,d}$ & $\varphi_l$ & $\phi_l$ & $\chi^0$ & $\varphi_\nu$& $\phi_\nu$& $\psi^0$ & $\omega^0$ \\ \hline
$S4$ &$3_1$ &$1_1$ & $2$ & $1_1$ & $1_2$ & $1_1$ & $3_1$ & $3_2$ & $3_1$ & $3_1$ & $3_2$ & $1_1$ &$2$ \\
$Z3$ & 1 & $\omega^2$ & $\omega^2$& 1 & 1 & 1 &  $\omega$& $\omega$ &$\omega$ & 1 & 1 & 1  & 1 \\
$U(1)_R$ & 0 & 1 & 1 & 1 &  1&  1& 0 & 0  & 2 & 0 & 0 & 2 & 2  \\ \hline
\end{tabular}
\caption{Transformation properties of all the fields under $S4$ $\times$ $Z3$ $\times$ $U(1)_R$.}
\end{table}

The transformation properties of the matter fields in the lepton sector and flavon fields needed are presented in Table 1. The 3 generations of left-handed lepton doublets are assigned to transform as $3_1$ under $S4$, while $e^c$ and ($\mu^c$,$\tau^c$) transform as $1_1$ and $2$ respectively. The two right-handed neutrinos $N_1$ and $N_2$ which are introduced to realize seesaw mechanism transform as $1_1$ and $1_2$ respectively. The $S4$ symmetry is then spontaneously broken by flavon fields, singlets under SM gauge group. The subindex of flavon fields $\varphi_l$, $\phi_l$, $\varphi_\nu$ and $\phi_\nu$ indicate their roles in generating charged lepton and neutrino masses respectively. Since we discuss the model in the context of supersymmetry, two Higgs doublets which are singlets under $S4$ are included. A $U(1)_R$ symmetry is also introduced which will play an important role in discussing the VEV alignment as we shall see in Section 4. In addition, a $Z3$ symmetry is added which can significantly constrain the form of Yukawa coupling and scalar potential as shown in the following. As far as the quark sector is concerned, we naively assume that quark fields transform as singlets under $S4\times Z3$ and get masses via conventional means, so we only discuss the lepton sector in the present paper.

The complete superpotential can be written as $W$= $W_l$ + $W_\nu$ +$W_d$,
\begin{equation}
W_l=y_1e^cLh_d\varphi_l/\Lambda+ y_2(\mu^c,\tau^c)Lh_d\varphi_l/\Lambda
+ y_3(\mu^c,\tau^c)Lh_d\phi_l/\Lambda ,
\end{equation}
\begin{equation}
W_\nu=y_4N_1Lh_u\varphi_\nu/\Lambda+y_5N_2Lh_u\phi_\nu/\Lambda+M_1N_1N_1+M_2N_2N_2 ,
\end{equation}
where $\Lambda$ is the cut-off scale of the theory and $W_d$ which describes the scalar potential will be given in Section 5. As mentioned above, the $Z3$ symmetry constrains $\varphi_l$ and $\phi_l$ to only couple with right-handed charged leptons so that give the charged lepton masses, highlighting the meaning of the subindex; so are $\varphi_\nu$ and $\phi_\nu$.

We assume that flavon fields get the following VEV configurations \cite{17} for the time being and then show it is a natural result in Section 4:

\begin{equation}
    \langle\varphi_l\rangle=\left(\begin{array}{c}
     A \\
      0\\
      0
\end{array}
\right)\hspace{2cm}
\langle\phi_l\rangle=\left(\begin{array}{c}
     B \\
      0\\
      0
\end{array}
\right),
\end{equation}

\begin{equation}
     \langle\varphi_\nu\rangle=\left(\begin{array}{c}
    0\\
     -C\\
      C
\end{array}
\right)\hspace{2cm}
\langle\phi_\nu\rangle=\left(\begin{array}{c}
     D \\
      D\\
      D
\end{array}
\right),
\end{equation}
where $A$, $B$, $C$, $D$ should obey the following relations,

\begin{equation}
   g_1 A^2+ g_3 B^2=0,
\end{equation}

\begin{equation}
  C^2=\frac{g_8}{2g_4 g_8 + g_5 g_6}M_{\psi^0} \hspace{2cm}
  D^2=-\frac{g_6}{6g_4 g_8 +3g_5 g_6}M_{\psi^0}.
\end{equation}

With the above specific VEV configuration, the charged lepton mass matrix is diagonal when the Higgs field also gets VEV:

\begin{equation}
     M_l=V_d \left(\begin{array}{ccc}
     y_1A/\Lambda&0 &0\\
       0&y_2A/\Lambda+y_3B/\Lambda&0\\
       0&0&y_2A/\Lambda-y_3B/\Lambda
\end{array}
\right),
\end{equation}
where $V_d$ is the VEV of $h_d$. Referring to the experimental data on the charged lepton masses, $y_1V_dA/\Lambda$, $y_2V_dA/\Lambda$ and $y_3V_dB/\Lambda$ are about $0.5 MeV$, $900 MeV$ and $-800 MeV$ respectively. If we assume that all Yukawa coupling constants and all VEVs of flavon fields are both of the same order of magnitude, the hierarchy between $0.5 MeV$ and $900 MeV$ and $-800 MeV$ poses a problem. In order to explain the hierarchy, we can take into consideration the well-known Froggatt-Nielsen mechanism \cite{18} and only need to assign $e^c$ a $F-N$ number $N$, so that electron mass becomes $y_1V_dt^NA/\Lambda$ with $t$, a small number, characterizing the ratio of the VEV of Froggatt-Nielsen flavon field to the masses of heavy Froggatt-Nielsen Fermion fields. From now on, we will not write Froggatt-Nielsen flavon field explicitly and just need to keep in mind that terms involving $e^c$ have an additional suppression of $t^N$, of order $10^{-3}$. If we parameterize $\langle\Phi\rangle$/$\Lambda$ as $u$ where $\langle\Phi\rangle$ represents the common magnitude of flavon fields VEVs, we can get a lower bound for $u$ from the requirement for Yukawa couplings to be perturbative. In the case of small $\tan \beta$, we can obtain $u$ $\sim$ $0.01$ from $yV_d\langle\Phi\rangle/\Lambda\sim1 GeV$ when $y$ reaches order 1.

The mass matrices for right-handed Majorana and Dirac neutrinos \cite{9} are

\begin{equation}
     M_N=\left(\begin{array}{cc}
     M_1 &0 \\
       0&M_2
\end{array}
\right)\hspace{2cm}
 M_D=V_u \left(\begin{array}{cc}
          0 &y_5D/\Lambda\\
      y_4C/\Lambda&y_5D/\Lambda\\
       -y_4C/\Lambda&y_5D/\Lambda
\end{array}
\right),
\end{equation}
where $V_u$ is the VEV of $h_u$.

For convenience, we denote $y_4V_uC/\Lambda$ and $y_5V_uD/\Lambda$ as $a$ and $b$. After seesaw, the light Majorana neutrino mass matrix is given by,
\begin{spacing}{1.5}
\begin{equation}
M_\nu=M_D M^{-1}_N M^T_D=\left(\begin{array}{ccc}
\frac{b^2}{M_2}&\frac{b^2}{M_2}&\frac{b^2}{M_2}\\
\frac{b^2}{M_2}&\frac{b^2}{M_2}+\frac{a^2}{M_1}&\frac{b^2}{M_2}-\frac{a^2}{M_1}\\
\frac{b^2}{M_2}&\frac{b^2}{M_2}-\frac{a^2}{M_1}&\frac{b^2}{M_2}+\frac{a^2}{M_1}
\end{array}
\right),
\end{equation}
\end{spacing}
which is diagonalized by the tri-bimaximal mixing matrix with three eigenvalues being $m_1$=0, $m_2$=$\frac{3b^2}{M_2}$ and $m_3$=$\frac{2a^2}{M_1}$. In this case, $m_2$=$\sqrt{\Delta m^2_{12}}$ $\sim$ $0.009$ $eV$, $m_3$=$\sqrt{\Delta m^2_{13}}$ $\sim$ $0.049$ $eV$, resulting in $\sum m_i$ $\sim$ $0.058$ $eV$. Unfortunately, $|m_{ee}|$ is about $0.003 eV$, far below the sensitivity of present and future experiments \cite{19} searching for neutrino-less double beta decay. As pointed out in \cite{11}, this model cannot fulfill the role of leptogenesis \cite{20}, due to the vanishing of $Im (M_D M^\dag_D)_{12}$. As analyzed in the charged lepton sector, $yV_u\langle\Phi\rangle/\Lambda$ $\sim$ $1 GeV$ in the case of low $\tan \beta$, so $M_1$ and $M_2$ are of the order of $10^{11}$ $GeV$ about lower than in the conventional seesaw mechanism by $4$ orders of magnitude.

\section{The VEV alignment}

In this section, we show how to get the specific VEV alignment used in the above section. In table $1$, there are several driving fields denoted by subindex $'0'$ whose $U(1)_R$ charge is $2$. The driving fields are gauge singlets and assumed to achieve no VEV. Due to the $U(1)_R$ symmetry, the superpotential terms involving driving fields has to take on the form of one driving field multiplied by combinations of flavon fields. In addition, the $Z3$ symmetry constrains $\varphi_l$ and $\phi_l$ to couple with $\chi^0$, and $\varphi_\nu$ and $\phi_\nu$ to couple with $\psi^0$ and $\omega^0$. Consequently, the driving superpotential can be written as,
\[
W_d=g_1\chi^0(\varphi_l\varphi_l)+ g_2\chi^0(\varphi_l\phi_l)+g_3\chi^0(\phi_l\phi_l)
+M_{\psi^0}\psi^0+g_4\psi^0(\varphi_\nu\varphi_\nu)
\]
\[
+g_5\psi^0(\phi_\nu\phi_\nu)+g_6\omega^0(\varphi_\nu\varphi_\nu)+ g_7\omega^0(\varphi_\nu\phi_\nu)+ g_8\omega^0(\phi_\nu\phi_\nu).
\]

The conditions for the driving fields not to get VEVs can be obtained by deriving $W_d$ with respect to each driving field,

\begin{eqnarray}
\left\{
\begin{array}{lll}
g_1(2\varphi_{l1}^2-2\varphi_{l2}\varphi_{l3})+g_2(\varphi_{l2}\phi_{l3}-\varphi_{l3}\phi_{l2})+g_3(2\phi_{l1}^2-2\phi_{l2}\phi_{l3})&=& 0 \\
 g_1(2\varphi_{l2}^2-2\varphi_{l1}\varphi_{l3})+g_2(\varphi_{l3}\phi_{l1}-\varphi_{l1}\phi_{l3})+g_3(2\phi_{l2}^2-2\phi_{l1}\phi_{l3})&=& 0\\
 g_1(2\varphi_{l3}^2-2\varphi_{l1}\varphi_{l2})+g_2(\varphi_{l1}\phi_{l2}-\varphi_{l2}\phi_{l1})+g_3(2\phi_{l3}^2-2\phi_{l1}\phi_{l2})&=&0
\end{array},
\right.
\end{eqnarray}

\begin{eqnarray}
\left\{
\begin{array}{lll}
M_{\psi^0}+g_4(\varphi_{\nu1}^2+2\varphi_{\nu2}\varphi_{\nu3})+g_5(\phi_{\nu1}^2+2\phi_{\nu2}\phi_{\nu3})&=& 0 \\
 g_6(\varphi_{\nu3}^2+2\varphi_{\nu2}\varphi_{\nu1})-g_7(\varphi_{\nu3}\phi_{\nu3}+\varphi_{\nu1}\phi_{\nu2}+\varphi_{\nu2}\phi_{\nu1})
 +g_8(\phi_{\nu3}^2+2\phi_{\nu1}\phi_{\nu2})&=& 0\\
 g_6(\varphi_{\nu2}^2+2\varphi_{\nu1}\varphi_{\nu3})+g_7(\varphi_{\nu2}\phi_{\nu2}+\varphi_{\nu1}\phi_{\nu3}+\varphi_{\nu3}\phi_{\nu1})
 +g_8(\phi_{\nu2}^2+2\phi_{\nu1}\phi_{\nu3})&=& 0
\end{array}.
\right.
\end{eqnarray}

The three equations in Eq.(11) are satisfied by the VEV alignments for $\varphi_l$ and $\phi_l$ in Eq.(4) with $A$ and $B$ satisfying Eq.(6). However, the flat direction described by Eq.(6) poses a problem because we can only define the relation between $A$ and $B$ rather than determine their concrete values. If we include soft mass terms $m^2_{\varphi_l}$ and $m^2_{\phi_l}$ and assume that they are negative, $\langle \varphi_l \rangle$ can slide to a large scale and then values of $A$ and $B$ can be fixed, as argued by Altarelli et al. in \cite{15}. The three equations in Eq.(12) are satisfied by the VEV alignments for $\varphi_\nu$ and $\phi_\nu$ in Eq.(5) and $C^2$ and $D^2$ can be determined, given in Eq.(7). In a word, the VEV alignment used in this paper is the minimum of the scalar potential.

\section{NLO corrections}

In this section, we firstly analyze the next-to-leading-order(NLO) corrections to $W_d$ which lead to deviations of flavon VEVs from the LO results presented in Eqs.(4-5) and then discuss the NLO corrections to the mass matrices for charged leptons and neutrinos. Finally, we formally present how the mixing matrix is corrected up to the order of $u$. As a matter of fact, all possible $S4 \times Z3$ invariant terms of dimension-six should be included at NLO. The NLO operators will be suppressed by a factor $\langle\Phi\rangle$/$\Lambda$ $\sim$ $u$ with respect to the LO operators.

$W_d$ can now be written as $W_d$ = $W^0_d$ + $\Delta W_d$, where $\Delta W_d$ are all possible dimension-6 and $S4 \times Z3$ invariant terms involving driving fields. Terms contained in $\Delta W_d$ can be obtained by inserting one flavon field to the LO driving superpotential and are suppressed by $\Lambda$. Because of the $Z3$ symmetry, only insertion of $\varphi_\nu$ or $\phi_\nu$ is allowed. Hence, $\Delta W_d$ can be written as following,

\begin{equation}
\Delta W_d=\frac{1}{\Lambda}(\sum r_i I^{\chi^0}_i + \sum s_i I^{\psi^0}_i +\sum t_i I^{\omega^0}_i)
\end{equation}
where $r_i$, $s_i$ and $t_i$ are coefficients, $I^{\chi^0}_i$, $I^{\psi^0}_i$ and $I^{\omega^0}_i$ are NLO operators containing $\chi^0$, $\psi^0$ and $\omega^0$ respectively. The complete terms are explicitly listed below(we indicate the representation R with $(\cdots)_R$):

\begin{tabular}{lll}
$I^{\chi^0}_1=\chi^0[(\varphi_l \varphi_l)_1 \varphi_\nu]$& $I^{\chi^0}_2=\chi^0[(\varphi_l \varphi_l)_2 \varphi_\nu]$&$I^{\chi^0}_3=\chi^0[(\varphi_l \varphi_l)_2 \phi_\nu]$\\
$I^{\chi^0}_4=\chi^0[(\varphi_l \varphi_l)_{3_1} \varphi_\nu]$&$I^{\chi^0}_5=\chi^0[(\varphi_l \varphi_l)_{3_1} \phi_\nu]$&$I^{\chi^0}_6=\chi^0[(\phi_l \phi_l)_1 \varphi_\nu]$\\
$I^{\chi^0}_7=\chi^0[(\phi_l \phi_l)_2 \varphi_\nu]$&$I^{\chi^0}_8=\chi^0[(\phi_l \phi_l)_2 \phi_\nu]$&$I^{\chi^0}_9=\chi^0[(\phi_l \phi_l)_{3_1} \varphi_\nu]$\\

$I^{\chi^0}_{10}=\chi^0[(\phi_l \phi_l)_{3_1} \phi_\nu]$&$I^{\chi^0}_{11}=\chi^0[(\varphi_l \phi_l)_{1_2} \phi_\nu]$&$I^{\chi^0}_{12}=\chi^0[(\varphi_l \phi_l)_2 \varphi_\nu]$\\

$I^{\chi^0}_{13}=\chi^0[(\varphi_l \phi_l)_2 \phi_\nu]$&$ I^{\chi^0}_{14}=\chi^0[(\varphi_l \phi_l)_{3_1}  \varphi_\nu]$&$I^{\chi^0}_{15}=\chi^0[(\varphi_l \phi_l)_{3_1} \phi_\nu]$\\

$I^{\chi^0}_{16}=\chi^0[(\varphi_l \phi_l)_{3_2}  \varphi_\nu]$&$I^{\chi^0}_{17}=\chi^0[(\varphi_l \phi_l)_{3_2}  \phi_\nu]$&$I^{\psi^0}_1=\psi^0[(\varphi_\nu \varphi_\nu)_{3_1} \varphi_\nu]$\\

$I^{\psi^0}_2=\psi^0[(\varphi_\nu \phi_\nu)_{3_1} \varphi_\nu]$&$I^{\psi^0}_3=\psi^0[(\phi_\nu \phi_\nu)_{3_1} \varphi_\nu]$&$I^{\omega^0}_1=\omega^0[(\varphi_\nu \varphi_\nu)_{3_1} \varphi_\nu]$\\

$I^{\omega^0}_2=\omega^0[(\varphi_\nu \varphi_\nu)_{3_1} \phi_\nu]$&$I^{\omega^0}_3=\omega^0[(\phi_\nu \phi_\nu)_{3_1} \varphi_\nu]$&$I^{\omega^0}_4=\omega^0[(\phi_\nu \phi_\nu)_{3_1} \phi_\nu]$
\end{tabular}

The corrected VEV alignment is obtained by deriving the new superpotential with respect to the driving fields. However, the number of equations is smaller than the number of VEVs, we cannot solve the correction terms explicitly. Nevertheless, it can be seen that the VEV alignment given by Eqs.(4-7) is indeed a stable minimum and the correction terms are smaller than LO terms by a factor $u$ which is about $0.01$ when Yukawa coupling constants are of order $1$. In the following, we will omit the corrections to VEV alignments which will not affect the final result.

The NLO operators contributing to the charged lepton masses are obtained by inserting one $\varphi_\nu$ or $\phi_\nu$ to the LO Yukawa coupling superpotential and enumerated in the following,

\[
\Delta W_l=e^cLh_d[y^{(1)}_l(\varphi_l\varphi_\nu)+y^{(2)}_l(\varphi_l\phi_\nu)+y^{(3)}_l(\phi_l\varphi_\nu)+y^{(4)}_l(\phi_l\phi_\nu)]_{3_1}/\Lambda^2
\]
\[
+[(\mu^c,\tau^c)Lh_d]_{3_1}[y^{(5)}_l(\varphi_l\varphi_\nu)+y^{(6)}_l(\varphi_l\phi_\nu)+y^{(7)}_l(\phi_l\varphi_\nu)+y^{(8)}_l(\phi_l\phi_\nu)]_{3_1}/\Lambda^2
\]
\[
+[(\mu^c,\tau^c)Lh_d]_{3_2}[y^{(9)}_l(\varphi_l\varphi_\nu)+y^{(10)}_l(\varphi_l\phi_\nu)+y^{(11)}_l(\phi_l\varphi_\nu)+y^{(12)}_l(\phi_l\phi_\nu)]_{3_2}/\Lambda^2.
\]

Replacing the flavon fields and Higgs field with their LO VEVs in the terms above, we can obtain the corrected charged lepton mass matrix. For convenience, only the order of magnitude of mass matrix entries are given in the following,

\begin{equation}
     M_l=V_d \left(\begin{array}{ccc}
     \mathcal{O}(ut^N)& \mathcal{O}(u^2t^N) &\mathcal{O}(u^2t^N)\\
        \mathcal{O}(u^2)& \mathcal{O}(u/10)& \mathcal{O}(u^2)\\
       \mathcal{O}(u^2)&\mathcal{O}(u^2)&\mathcal{O}(u)
\end{array}
\right).
\end{equation}

Correspondingly, the unitary matrix that diagonalizes the mass matrix $M_lM^\dag_l$ at NLO can be written in the form:

\begin{equation}
     U_l=\left(\begin{array}{ccc}
     1& 0 &0\\
      0& 1& V^l_{23}u\\
      0&-V^l_{23}u&1
\end{array}
\right),
\end{equation}
where $V^l_{23}$ is coefficient of order one, $V^l_{12}$ and $V^l_{13}$ are of order $0.1$ and are omitted.

Similarly, the NLO operators contributing to neutrino mass matrix are also obtained by inserting one $\varphi_\nu$ or $\phi_\nu$ and given in the following,

\[
\Delta W_\nu=Lh_uN_1[y^{(1)}_\nu(\varphi_\nu\varphi_\nu)+y^{(2)}_\nu(\varphi_\nu\phi_\nu)+y^{(3)}_\nu(\phi_\nu\phi_\nu)]/\Lambda^2+Lh_uN_2[y^{(4)}_\nu(\varphi_\nu\phi_\nu)]/\Lambda^2
\]

The modified Dirac neutrino mass matrix can be obtained by replacing the scalar fields with their VEVs,

\begin{equation}
     \Delta M_D=V_u/\Lambda^2\left(\begin{array}{cc}
     2y^{(1)}_\nu C^2-2y^{(2)}_\nu CD& 0\\
        2y^{(1)}_\nu C^2+y^{(2)}_\nu CD& -3y^{(4)}_\nu CD\\
       2y^{(1)}_\nu C^2+y^{(2)}_\nu CD&3y^{(4)}_\nu CD
\end{array}
\right).
\end{equation}
For convenience, $\Delta M_D$ can be rewritten as,
\begin{equation}
     \Delta M_D=u\left(\begin{array}{cc}
          c& 0\\
        d& -e\\
       d&e
\end{array}
\right),
\end{equation}
where we denote $C/\Lambda$, $(2y^{(1)}_\nu C-2y^{(2)}_\nu D) V_u/\Lambda$, $(2y^{(1)}_\nu C+y^{(2)}_\nu D) V_u/\Lambda$ and \\$3y^{(4)}_\nu D V_u/\Lambda$ as $u, c, d$ and $e$ respectively.

The NLO correction to the right-handed neutrino masses is the following,
$$
\Delta M_N=f_1 N_1 N_1 \varphi_\nu \varphi_\nu / \Lambda + f_2 N_1 N_1 \phi_\nu \phi_\nu / \Lambda +  f_3 N_2 N_2 \varphi_\nu \varphi_\nu / \Lambda
$$
$$
+ f_4 N_2 N_2 \phi_\nu \phi_\nu / \Lambda + f_5 N_1 N_2 \varphi_\nu \phi_\nu / \Lambda.
$$
The last term vanishes when the scalar fields are replaced with their VEVs in Eq.(5), so there is no mixing term between $N_1$ and $N_2$ up to the NLO corrections; as a matter of fact, if we assume that $M_1$ and $M_2$ are of $\mathcal{O} (\Lambda)$, then terms in $\Delta M_N$ are suppressed by $u^2$ with respect to terms in $M_N$. Thus, we can disregard the NLO corrections to the right-handed neutrino masses.

The modified light neutrino Majorana mass matrix can be computed as follows,
\begin{spacing}{1.5}
\begin{equation}
     \Delta M_\nu=M_D M^{-1}_N \Delta M^T_D+\Delta M_D M^{-1}_N M^T_D
     \end{equation}
     \begin{equation}
     =u\left(\begin{array}{ccc}
     0&\frac{ac}{M_1}-\frac{be}{M_2}&-\frac{ac}{M_1}+\frac{be}{M_2}\\
     \frac{ac}{M_1}-\frac{be}{M_2}&\frac{2ad}{M_1}-\frac{2be}{M_2}&0\\
     -\frac{ac}{M_1}+\frac{be}{M_2}&0&-\frac{2ad}{M_1}+\frac{2be}{M_2}
     \end{array}
     \right).
\end{equation}
\end{spacing}

Resulting from the smallness of the NLO corrections, the unitary matrix that diagonalizes $M_\nu$ has a small deviation from the tri-bimaximal mixing. As a consequence, we can expand $U_\nu$ around $U_{TB}$ to the leading order of $u$,

\begin{equation}
U_\nu=U_{TB}+\Delta U_\nu u ,
\end{equation}
where $\Delta U_\nu$ are parameterized by three parameters $V^\nu_{12}$, $V^\nu_{13}$ and $V^\nu_{23}$,
\begin{equation}
\Delta U_\nu=\left(\begin{array}{ccc}
     -\sqrt{\frac{1}{3}}V^\nu_{12}& \sqrt{\frac{2}{3}}V^\nu_{12} &V^\nu_{13}\\
     -\sqrt{\frac{1}{3}}V^\nu_{13}&-\sqrt{\frac{1}{6}}V^\nu_{13}&  \sqrt{\frac{1}{2}}V^\nu_{23}\\
      -\sqrt{\frac{1}{3}}V^\nu_{13}&-\sqrt{\frac{1}{6}}V^\nu_{13}&-\sqrt{\frac{1}{2}}V^\nu_{23}
\end{array}
\right).
\end{equation}

Finally, the mixing matrix in the lepton sector can be obtained through the following relation,

\begin{equation}
U=U^\dag_l U_\nu.
\end{equation}
As a result,
 \begin{equation}
 \sin{\theta_{23}} \sim \frac{1}{\sqrt{2}} + \frac{1}{\sqrt{2}}V^\nu_{23}u-V^l_{23}u,
  \end{equation}
  \begin{equation}
  \sin{\theta_{13}}\sim V^\nu_{13}u,
  \end{equation}
 \begin{equation}
 \sin{\theta_{12}} \sim \frac{1}{\sqrt{3}} + \sqrt{\frac{2}{3}}V^\nu_{12}u.
  \end{equation}
 We can see that the NLO corrections lead to deviations of order $u$ from the tri-bimaximal mixing.

\section{Conclusions and discussions}
In conclusion, with $S4$ as the family symmetry, we manage to obtain the tri-bimaximal mixing in the lepton sector in the context of minimal seesaw in which only two right-handed neutrinos are introduced. In order to prevent unwanted terms, an additional $Z3$ symmetry is introduced which plays an important role in the model as we have seen. In the model, the fields are assigned to appropriate transformation properties under $S4 \times Z3$. When the flavon fields get particular VEV configurations, the charged leptons get a diagonal mass matrix in the flavor basis while the mass matrix in the neutrino sector holds a special form which is diagonalized by the right tri-bimaximal matrix. The success of the model heavily relies on the specific transformation properties of different fields under family symmetry and the special forms of the VEVs of flavon fields. Accordingly, whether the particular VEV configuration used in the model can be got naturally is a key question. Fortunately, the VEV configuration proves to be the minimum of scalar potential as we see in Section $4$. NLO corrections have also been considered, and make the mixing matrix deviate from the tri-bimaximal pattern by quantities proportional to $u$ whose value is about $0.01$. Phenomenologically, the mass spectrum is of normal hierarchy with $m_1=0$, and $\sum m_i$ and $|m_{ee}|$ are predicted to be about $0.058$ $eV$ and $0.003$ $eV$ respectively. Although the phenomenological consequences are not rich, the predictions are definite, distinguishing the model from others.

\section{Acknowledgements}
We would like to thank Chun Liu and Hua Shao for helpful discussions. This work was supported in part by the National Natural Science Foundation of China under Nos. 11075193 and 10821504, and by the National Basic Research Program of China under Grant No. 2010CB833000.

\end{document}